\documentclass[superscriptaddress,aps,showpacs,nofootinbib,
preprintnumbers,twocolumn]{revtex4}
\usepackage{amsmath,amssymb}
\usepackage{graphicx}% Include figure files
\usepackage{dcolumn}% Align table columns on decimal point
\newcommand{\mgsq}{M^2}

\newcommand{\beq}{\begin{equation}}
\newcommand{\eeq}{\end{equation}}
\newcommand{\beqa}{\begin{eqnarray}}
\newcommand{\eeqa}{\end{eqnarray}}
\newcommand{\beqar}{\begin{eqnarray*}}
\newcommand{\eeqar}{\end{eqnarray*}}

\begin{document}

\preprint{UG-FT-210/06, CAFPE-79/06, ROMA-1439-06}

\title{
New physics from ultrahigh energy cosmic rays
}% Force line breaks with \\

%\textbf{\Large
%New physics from ultrahigh energy cosmic rays}
% Force line breaks with \\
%\vspace{40pt}

\author{J. I. Illana}
\email{jillana@ugr.es}
\author{M. Masip}
\email{masip@ugr.es}
\affiliation{
CAFPE and
Depto.~de F{\'\i}sica Te\'orica y del Cosmos, Universidad de Granada,
18071 Granada, Spain}

\author{D. Meloni}
\email{meloni@roma1.infn.it}
\affiliation{
INFN and 
Dipto.~di Fisica, Universit\`a degli Studi di Roma ``La Sapienza", 
00185 Rome, Italy}

%\centerline{\today}% It is always \today, today,

\begin{abstract}
Cosmic rays from outer space enter the atmosphere with energies
of up to $10^{11}$~GeV. The initial particle or a secondary
hadron inside the shower may then interact 
with an air nucleon
to produce nonstandard particles. In this article we study the
production of new physics by high energy cosmic rays, focusing
on the long-lived gluino of split-SUSY models and a WIMP
working as dark matter.
We first deduce the total flux of hadron events at any
depth in the atmosphere, showing that secondary hadrons can not
be neglected. Then we use these results to find the flux of gluinos
and WIMPs that reach the ground after being produced inside air
showers. We also evaluate the probability of producing these 
exotic particles in a single proton shower of ultrahigh energy. 
Finally we discuss the possible signal in
current and projected experiments. While the tiny flux of 
WIMPs does
not seem to have any phenomenological consequences, 
we show that the gluinos could modify substantially the profile 
of a small fraction
of extensive air showers. In particular, they could produce 
a distinct signal observable at AUGER in showers of large zenith 
angle.

\end{abstract}

%\keywords{Neutrino Masses, Higgs Physics, Beyond Standard Model}
                              %Use showkeys class option if keyword
\maketitle

\section{Introduction}

The standard model (SM) has been very succesful when confronted
with the experimental data from particle colliders. 
Despite that, not much is
known about the nature of the Higgs sector. Hopefully, 
the Large Hadron Collider (LHC) at CERN will provide the value of
the Higgs mass and will unveil the mechanism that explains 
that value. A dynamical mechanism would imply new physics just
above the electroweak scale: SUSY, little Higgs, technicolor, 
extra dimensions, or some other unexpected physics.
It is also possible, however, that the LHC just completes
and confirms the SM as it was formulated, pushing above the TeV
scale the limits for this new physics 
\cite{Agrawal:1997gf,Arkani-Hamed:2004fb}. This 
possibility has been seriously considered 
after recent astrophysical and cosmological data 
suggested a non-zero vacuum energy density \cite{Bennett:2003bz}.
The preferred value $\Lambda\approx (10^{-3}\;{\rm eV})^4$ 
does not seem to be explained
by any dynamics at that scale, and could be just telling us
that the universe is {\it larger} and/or {\it older}
than we expected\footnote{$\Lambda$ could take different values 
in different {\it regions}. To expect one region
(the universe we see) with a value $10^{120}$ times smaller than 
its natural value, around $M_{Planck}^4$, there should be over 
$10^{120}$ of these regions. 
Analogously, our universe could have been preceded by 
$10^{120}$ Big Bang/Big Crunches of different vacuum energy.}: 
a multiverse \cite{Weinberg:1987dv}. 
From the experimental point of view, precision data 
(flavor physics, EDMs) \cite{Erler:2004nh} suggest 
that the scale of new physics is well
above the TeV, whereas neutrino oscillations point to an even larger 
scale, above $10^{11}$~GeV \cite{seesaw}. In contrast, dark matter 
strongly suggests a stable WIMP below (or at) the TeV \cite{Lee:1977ua}.
Therefore, the question of 
what (if any) physics beyond the SM 
we should expect at accessible energies is wide open.

On the other hand, it is known that accelerators are not the
only place to look for collisions of TeV energy. There is a well
established spectrum of cosmic rays that extends up to $E=10^{11}$~GeV
\cite{Yao:2006px}.
When a proton from outer space enters the atmosphere and hits a nucleon, 
the center of
mass energy $\sqrt{s}=\sqrt{2m_N E}$ in the collision may go up to 
500 TeV. Depending on the energy of the initial particle, after the 
first interaction there will be many 
secondary hadrons \cite{Greisen:1960wc,Lipari:1993hd}
with still enough energy to produce TeV physics in a collision
with an atmospheric nucleon. If created, the massive particle 
would be {\it inside} the air shower together with thousands of other 
standard particles. 
Therefore, to have any chance to detect it the new particle should
be long lived (its decay products inside the shower would be 
undetectable). This is precisely the case of the gluino of split-SUSY 
models \cite{Arkani-Hamed:2004fb}. Such a particle could 
change substantially the profile of the air shower where it was 
produced or it may  {\it survive} (together with muons and neutrinos)
the shower and be observed far from the initial interaction point, in 
quasi-horizontal events or in neutrino telescopes.

In this paper we study the production
of new physics by cosmic rays. To evaluate that we determine the 
total flux of hadrons in the atmosphere: 
primary plus secondary nucleons, pions and kaons. 
The decay length of these particles 
at these energies is much larger than their interaction length
in air, so the probability that they produce new physics
is just
\beq
{\cal P}_{X} =
{\sigma_{X}\over \sigma_{T}}\;,
\label{eq1}
\eeq
where $\sigma_{X}$ is the cross section to produce the exotic
particle and $\sigma_{T}$ the total cross section
of the hadron with the air. We deduce as well the spectrum of
secondary hadrons created by a single proton of ultrahigh energy.
We then show how to apply these results to evaluate the 
production rate of long-lived gluinos and also of a WIMP working 
as dark matter of the universe, 
and discuss possible signals in air showers and neutrino telescopes. 

Other studies of the production of new physics by cosmic rays 
refer to primary neutrinos \cite{domokos}. 
In particular, there are several recent analyses of 
the possibility to observe long-lived {\it staus} at
neutrino telescopes \cite{Albuquerque:2003mi}.
A neutrino of very high energy 
may reach a telescope and interact there 
with a nucleon. Since the neutrino is 
weakly interacting, the relative effect of new physics is here
more important than for protons or electrons. 
If IceCube \cite{icecube} establishes
the neutrino flux at these energies, the search for deviations
to the standard neutrino-nucleon cross sections will put strong 
constraints on some types of TeV physics \cite{Anchordoqui:2005ey}.

\section{Total flux of hadrons}

\begin{figure}
\centerline{\includegraphics[width=\linewidth]{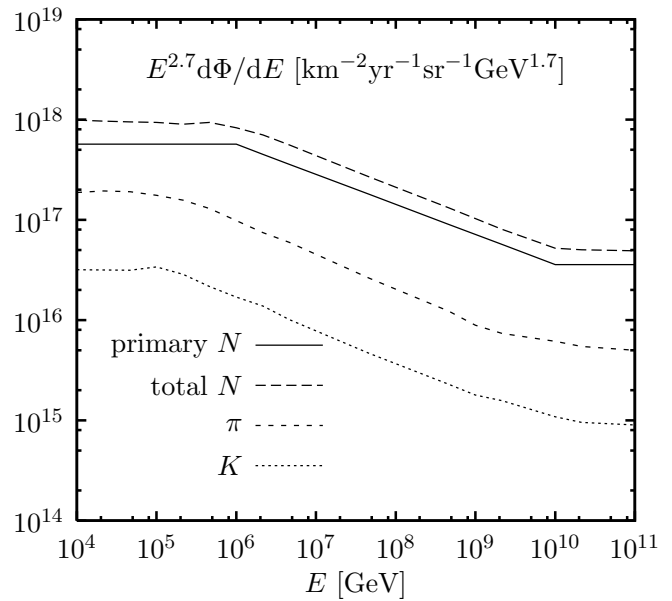}}
\caption{
Flux of primary nucleons (solid), total nucleons, and 
secondary pions and kaons. \label{fig1}}
\end{figure}

The flux of nucleons reaching the atmosphere in the range of energy
from several GeV to $\approx 10^6$~GeV can be approximated 
by \cite{Yao:2006px}
\beq
\frac{{\rm d}\Phi_{N}}{{\rm d}E}\approx 
1.8 \; E^{-2.7}\; {\rm {nucleons\over cm^2\;s\;sr\;GeV}}\;,
\label{eq2}
\eeq
with $E$ given in GeV. Around 90\% of these primary nucleons are
protons (free or bound in nuclei) and 10\% neutrons. At
energies $\approx 10^6$~GeV (the {\it knee}) the differential
spectral index changes from 2.7 to 3, and at $10^{10}$~GeV
(the {\it ankle}) it goes back to 2.7 (see Fig.~\ref{fig1}).

%%%%%%%%%%%%%%%%%%%%%%%%%%%%%%%%%%%%%%%%%%%%%%%%%%%%%%%%%%%%%%%%%%%%%%%%%%%%%%%
%%%%%%%%%%%%%%%%%%%%%%%%%%%%%%%%%%%%%%%%%%%%%%%%%%%%%%%%%%%%%%%%%%%%%%%%%%%%%%%

As this flux of nucleons enters the atmosphere, its 
interactions with the air will induce a flux of secondary
and tertiary hadrons (all of them referred as secondary)
at different vertical depths. Although approximate
analytic expressions can be obtained in limited regions of
energy \cite{Greisen:1960wc,Lipari:1993hd}, it is easy as 
well to perform a numerical simulation. 

We use {\tt CORSIKA} \cite{Heck:1998vt} to obtain the number 
and energy distribution of secondary nucleons, (charged) pions
and kaons (we neglect the baryons containing an $s$ quark) produced 
inside the air shower started by a nucleon of energy $E$.
In Fig.~\ref{fig1} we plot the total 
(primary plus secondary) flux of hadrons with energies from
$10^4$ to $10^{11}$~GeV. We obtain this flux generating 1500 showers
per decade of energy. We observe that at energies around
$10^7$~GeV ($\sqrt{s}\approx  5$ TeV) the secondary nucleons
increase in a 50\% the number of primaries. Note also that
the number of mesons (all of them secondary) is approximately 
a 15\% the number of primary nucleons of the same energy.

The region of energies beyond $10^8$~GeV (the one relevant at
AUGER \cite{auger}) deserves a more detailed analysis. We notice 
an important
factor that distinguishes two types of showers: the
presence or not of a {\it leading hadron} carrying a significant
fraction of energy after each hadronic interaction.
In Table~\ref{tab1} we give two examples of showers, both of energy
$E= 10^{10}$~GeV, with a very different spectrum of secondary
hadrons. Clearly, shower {\it (b)} does not include a leading
hadron after the first interaction. A relevant parameter to 
characterize this feature is the {\it elasticity} $x_F$ of 
the first hadronic interaction: in example {\it (a)} we have
$x_F\approx0.3$ ({\it i.e.,} a secondary hadron keeps 30\% of the 
initial energy) whereas {\it (b)} corresponds to $x_F\approx0.03$.
The analysis of 500 showers of $10^{10}$~GeV shows that around 
$15 \%$ of them have 
$x_F<0.06$ ({\it i.e.}, no leading
hadron after the first interaction with the air).

%%%%%%%%%%%%%%%%%%%%%%%%%%%%%%%%%%%%%%%%%%%%%%%%%%%%%%%%%%%%%%%%%%%%%%%%%%%%%%%
\begin{table}
\begin{center}
\begin{tabular}{|c|r|r|r||r|r|r|}
\hline
{Energy [GeV]} & $N\;\;$ & $\pi\;\;$ & $K\;\;$ & $N\;\;$ & $\pi\;\;$ & $K\;\;$ 
\\
\hline
$10^4$--$2.1\times 10^{4}$       & 7539 & 54585 & 10092 & 9873 & 70692 & 13026
\\
$2.1\times 10^{4}$--$4.6\times 10^{4}$ & 3907 & 27558 &  5196 & 4802 & 35700 &  6757
\\
$4.6\times 10^{4}$--$10^5$       & 2015 & 13628 &  2524 & 2476 & 17644 &  3360
\\
$10^5$--$2.1\times 10^{5}$       &  909 &  6771 &  1323 & 1269 &  8415 &  1634
\\
$2.1\times 10^{5}$--$4.6\times 10^{5}$ &  469 &  3352 &   608 &  600 &  4314 &   817
\\
$4.6\times 10^{5}$--$10^6$       &  254 &  1695 &   318 &  300 &  2065 &   390
\\
$10^6$--$2.1\times 10^{6}$       &  116 &   860 &   154 &  168 &  1010 &   200
\\
$2.1\times 10^{6}$--$4.6\times 10^{6}$ &   58 &   421 &    71 &   66 &   488 &    97
\\
$4.6\times 10^{6}$--$10^7$       &   24 &   197 &    32 &   23 &   230 &    39
\\
$10^7$--$2.1\times 10^{7}$       &   11 &    89 &    14 &   27 &   120 &    20
\\
$2.1\times 10^{7}$--$4.6\times 10^{7}$ &    6 &    36 &    14 &   13 &    60 &     9
\\
$4.6\times 10^{7}$--$10^8$       &    4 &    23 &     3 &    8  &   31 &    10
\\
$10^8$--$2.1\times 10^{8}$       &    0 &    17 &     2 &    2  &   16 &     2
\\
$2.1\times 10^{8}$--$4.6\times 10^{8}$ &    2 &     8 &     0 &    1 &     1 &     0
\\
$4.6\times 10^{8}$--$10^9$       &    0 &     1 &     2 &    0 &     0 &     0
\\
$10^9$--$2.1\times 10^{9}$       &    2 &     1 &     1 &    0 &     0 &     0
\\
$2.1\times 10^{9}$--$4.6\times 10^{9}$ &    1 &     0 &     0 &    0 &     0 &    0
\\
$4.6\times 10^{9}$--$10^{10}$    &    1 &     0 &     0 &    1 &     0 &     0
\\
\hline
\end{tabular}
\end{center}
\caption{%(a) elasticty=0.3042 [11], (b) elasticty=0.0286 [5]
Spectrum of hadrons in a $10^{10}$~GeV shower with a first interacion of
elasticity $x_F\approx0.3$ (left, shower {\it (a)} in the text) 
and $x_F\approx0.03$ (right, shower {\it (b)}). 
\label{tab1}}
\end{table}
%%%%%%%%%%%%%%%%%%%%%%%%%%%%%%%%%%%%%%%%%%%%%%%%%%%%%%%%%%%%%%%%%%%%%%%%%%%%%%%

\section{Flux of massive long-lived particles}

Let us now estimate the production rate of new physics 
in the collision of these cosmic hadrons with atmospheric 
nucleons. The probability 
that a hadron $h$
of energy $E$ produces the exotic particle(s) $X$ is 
determined by the cross section 
$\sigma^{hN}_X(E)$ and 
the total cross section with the air, 
$\sigma^{ha}_{T}(E)$: 
\beq
{\cal P}^h_X(E) \approx 
{A \; \sigma^{hN}_X \over \sigma^{ha}_{T}}\;,
\label{eq3}
\eeq
where we assume $A=14.6$ nucleons in a nucleus of air and neglect
nuclear effects in the interaction to produce $X$.
Taking the total fluxes ${\rm d}\Phi_h/{\rm d}E$ of the three species
of hadrons in Fig.~\ref{fig1} we
obtain that 
the flux of exotic particles (of any energy) is just
\beq
\Phi_{X} = \sum_{h=N,\pi,K}\; 
\int^{\infty}_{E_{\rm min}} {\rm d}E\;\frac{{\rm d}\Phi_{h}}{{\rm d}E}\; 
{\cal P}^h_X(E)\;.
\label{eq4}
\eeq
If we are interested in the energy distribution of $X$
we must evaluate 
\beq
\frac{{\rm d} \Phi_{X}}{{\rm d} E_X} =
\sum_{h}\; 
\int^{\infty}_{E_X} {\rm d}E\;\frac{{\rm d}\Phi_{h}}{{\rm d}E}\; 
\frac{A}{\sigma^{ha}_{T}}\; \frac{{\rm d}\sigma^{hN}_X} 
{{\rm d} E_X}
\;,
\label{eq5}
\eeq 
where $E_X$ is the energy of $X$ and the cross sections inside
the integral are evaluated at $E$. 
The (default) cross sections with the air used by {\tt CORSIKA}
above $10^4$~GeV can be approximated
by $\sigma^{ha}_{T}\approx C_0^h +
C_1^h \log (E/{\rm GeV})+C_2^h \log^2(E/{\rm GeV})$, 
with the constants given in Table~\ref{tab2}.

%%%%%%%%%%%%%%%%%%%%%%%%%%%%%%%%%%%%%%%%%%%%%%%%%%%%%%%%%%%%%%%%%%%%%%%%%%%%%%%
\begin{table}
\begin{center}
\begin{tabular}{|c|c|c|c|}
\hline
$h$ & $C_0^h$ [mb] & $C_1^h$ [mb] & $C_2^h$ [mb]
\\
\hline
$N$ & 185.7 & 13.3 & 0.08
\\
$\pi$ & 100.5 & 16.9 & 0.00
\\
$K$ & \;\;79.7 & 13.9 & 0.05
\\
\hline
\end{tabular}
\end{center}
\caption{
Constants defining the total cross section with the air. 
\label{tab2}}
\end{table}
%%%%%%%%%%%%%%%%%%%%%%%%%%%%%%%%%%%%%%%%%%%%%%%%%%%%%%%%%%%%%%%%%%%%%%%%%%%%%%%

%%%%%%%%%%%%%%%%%%%%%%%%%%%%%%%%%%%%%%%%%%%%%%%%%%%%%%%%%%%%%%%%%%%%%%%%%%%%%%%
\begin{figure}
\centerline{\includegraphics[width=\linewidth]{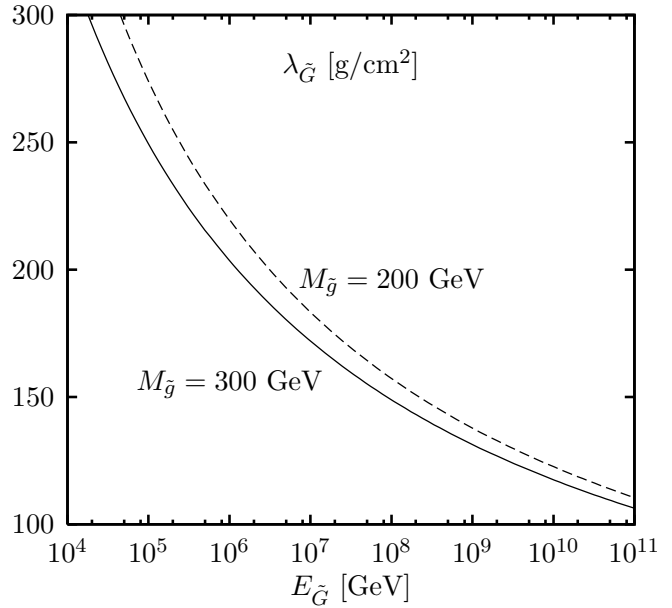}}
\caption{
Interaction length in air $\lambda_{\tilde G}$ of the gluino hadron. 
The interaction length in water is a 3.2\% shorter. \label{fig2}}
\end{figure}
%%%%%%%%%%%%%%%%%%%%%%%%%%%%%%%%%%%%%%%%%%%%%%%%%%%%%%%%%%%%%%%%%%%%%%%%%%%%%%%

As we mentioned above, to be of interest the
exotic particle should be {\it penetrating} and 
survive the air shower where it was produced. We will then
consider two different 
cases. The first one is the long-lived gluino 
$\tilde g$ of split-SUSY
models \cite{Arkani-Hamed:2004fb}. In the most {\it evasive}
scenario the gluino fragments {\it always} into an 
electrically neutral gluino-gluon hadronic
state $\tilde G$ (a particular type of {\it R-hadron}). 
The experimental bounds on its mass are in this case 
around 170 GeV \cite{Hewett:2004nw}. Notice that general
bounds \cite{Abazov:2006bj} or bounds based on the
observed delay in the time of flight versus a muon 
\cite{phillips} do not apply to a neutral long-lived 
hadron. The gluino hadron will interact often 
with the air, its interaction length (in Fig.~\ref{fig2})
can be estimated 
\cite{Baer:1998pg} as
$\lambda_{\tilde G} \approx (16/9)\;\lambda_\pi$ 
at $E_{\tilde G}\approx E_{\pi} M/m_{\pi}$ ({\it i.e.},
when both hadrons have the same velocity).
However, in each interaction 
it will lose a very small fraction of its 
energy $\Delta E/E \approx k/M$, where $k\approx $ 0.14--0.35
GeV and $M$ is the gluino mass 
\cite{Baer:1998pg,Kraan:2004tz,Hewett:2004nw}. 
Therefore, the gluino hadron is very penetrating. In particular,
it keeps going after the atmosphere has 
absorbed most of the hadrons in quasi-horizontal showers.  

%%%%%%%%%%%%%%%%%%%%%%%%%%%%%%%%%%%%%%%%%%%%%%%%%%%%%%%%%%%%%%%%%%%%%%%%%%%%%%%
\begin{figure}
\centerline{\includegraphics[width=\linewidth]{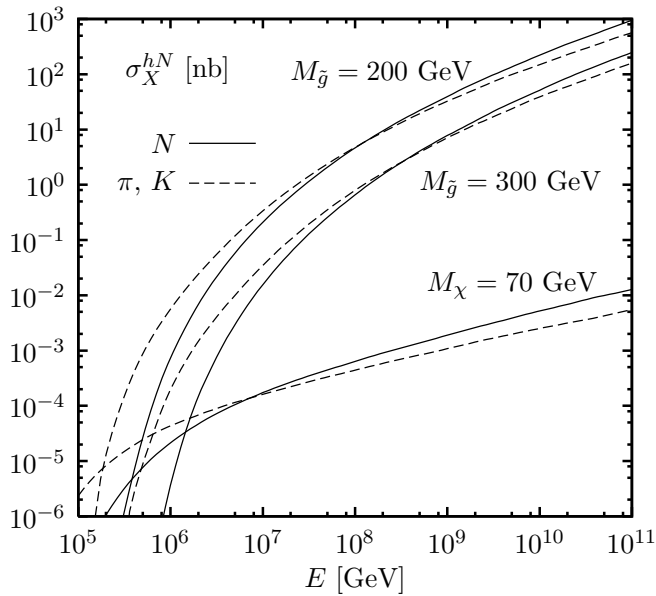}}
\caption{
Total cross sections $\sigma^{hN}_{X}$ with $X=\tilde g\tilde g$ and 
$\chi\chi$ for $h=N,\pi,K$. 
\label{fig3}}
\end{figure}
%%%%%%%%%%%%%%%%%%%%%%%%%%%%%%%%%%%%%%%%%%%%%%%%%%%%%%%%%%%%%%%%%%%%%%%%%%%%%%%

The parton-parton cross sections to produce gluino pairs
can be easily obtained taking the limit of large squark
mass in the general expressions given in Ref.~\cite{Dawson:1983fw}, 
and we include them in the Appendix. 
In Fig.~\ref{fig3} we plot the total $hN$ ($h=N,\pi,K$) cross 
sections $\sigma^{hN}_{\tilde g\tilde g}$ 
to produce gluino pairs for values of the 
hadron energy between $10^{4}$ and $10^{11}$~GeV
and $M=200,\;300$~GeV. We have assumed isospin
symmetry to deduce the kaon PDFs and have taken a renormalization
scale $\mu=0.2 M$, as suggested by a NLO calculation \cite{Beenakker:1996ch}.

%%%%%%%%%%%%%%%%%%%%%%%%%%%%%%%%%%%%%%%%%%%%%%%%%%%%%%%%%%%%%%%%%%%%%%%%%%%%%%%
\begin{figure}
\centerline{\includegraphics[width=\linewidth]{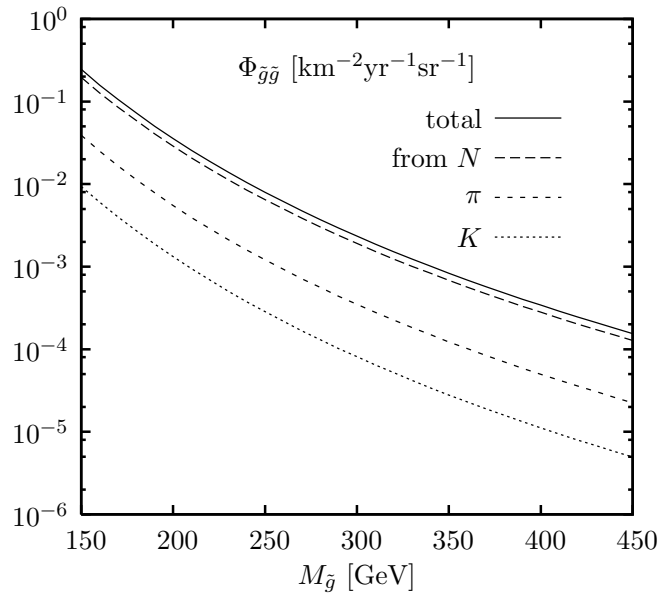}}
\caption{
Flux of gluino pairs as a function of the gluino mass. We separate
the contribution of the different hadrons to this flux\label{fig4}}
\end{figure}
%%%%%%%%%%%%%%%%%%%%%%%%%%%%%%%%%%%%%%%%%%%%%%%%%%%%%%%%%%%%%%%%%%%%%%%%%%%%%%%

\begin{figure}
\centerline{\includegraphics[width=\linewidth]{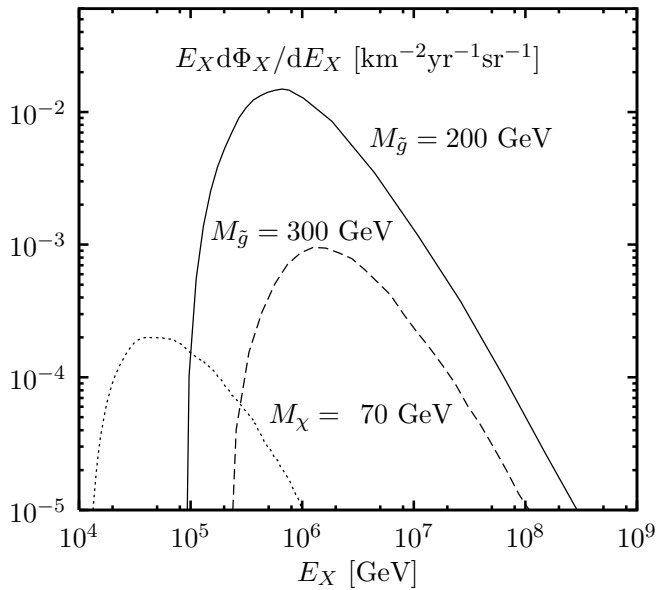}}
\caption{
Energy distribution of the gluino and WIMP events. \label{fig5}}
\end{figure}

We can now calculate the flux of gluino pairs produced in the
atmosphere. In Fig.~\ref{fig4} we plot the flux $\Phi_{\tilde g\tilde g}$ 
for values of $M$ between 150 and 450 GeV. We express this flux
in gluino pairs per year, squared kilometer and 
sterad. Multiplying by $2\pi$ sterad, for example, we can read that 
the $\approx \mbox{km}^2$ IceCube detector would be exposed to 
one downgoing gluino pair per year if $M\approx 160$~GeV.
Around a 64\% of the gluino flux would be produced by the primary 
nucleons, whereas the rest corresponds to secondary nucleons (16\%), 
pions (16\%) and kaons (4\%).
In Fig.~\ref{fig5} we give the energy distribution of the gluinos
for $M=200,\;300$~GeV. The energy $E$ is in this case 
the total energy of the two gluinos in the lab frame.
In the next section we comment on the phenomenological
relevance of these results.

%%%%%%%%%%%%%%%%%%%%%%%%%%%%%%%%%%%%%%%%%%%%%%%%%%%%%%%%%%%%%%%%%%%%%%%%%%%%%%%
%%%%%%%%%%%%%%%%%%%%%%%%%%%%%%%%%%%%%%%%%%%%%%%%%%%%%%%%%%%%%%%%%%%%%%%%%%%%%%%

The second nonstandard particle that we would like to consider
is a stable WIMP $\chi$ constituting the dark matter of the universe
(see \cite{Bertone:2004pz} for a recent review on dark matter).
To define a simple scenario, we assume that $\chi$ is a 
Majorana fermion with
only weak interactions and a relatively light mass, 
$M_\chi\approx 70$~GeV, and that all other possible 
particles (neutral or charged in the same $SU(2)_L$ 
multiplet) are 
at least 30 GeV heavier. The later requirement 
is necessary to avoid collider bounds, and will imply that 
{\it coannihilations} at temperatures below $M_\chi$ are
irrelevant. We also need a discrete symmetry similar
to the R-parity to make $\chi$ stable.
Under these hypotheses, the only relevant parameter
to determine the relic abundance of this particle is its
coupling to the $Z$ boson. We will assume a 
coupling $(g/2c_W)\epsilon$. The parameter $\epsilon$ fixes 
the annihilation rate of $\chi$ pairs into quarks and leptons 
(with a $Z$ in the $s$ channel)
and then the contribution of $\chi$ to the energy density of
the universe. 
It is a simple exercise to obtain that for
$M_\chi=70$~GeV we must have $\epsilon\approx 0.25$.
Ref.~\cite{Masip:2005fv} provides an explicit realization
of this model in a partly SUSY scenario.

We can now estimate the flux of $\chi$ pairs produced by
cosmic rays. Again, the $\sigma^{hN}_{\chi\chi}$ cross section 
\cite{Ma:1989jp} (see the Appendix) 
depends only on the coupling with the $Z$ boson (we neglect 
virtual effects of heavier particles
as well as the production of heavier fields that 
may decay into $\chi$ plus standard particles).
In Fig.~\ref{fig3} we have included 
$\sigma^{hN}_{\chi\chi}$. The flux of downgoing
dark-matter particles would be $4.6\times10^{-4}$ pairs per year, squared 
kilometer and sterad. Their energy distribution is shown in 
Fig.~\ref{fig5}. 

\section{New physics in a single event of ultrahigh energy}

We would also like to find the probability for the production
of gluino or WIMP pairs in single events of the highest 
energy (the ones to be measured at AUGER). In Table~\ref{tab1} we
give two simulations of 
showers of high ($x_F\approx 0.3$) and low ($x_F\approx 0.03$) elasticity, 
both with a proton primary of energy $10^{10}$~GeV.
For $M=200\;(300)$~GeV
the probability to produce a gluino pair in each shower is 
${\cal P}^a_{\tilde g\tilde g} \approx 5.4\times10^{-5}\;(9.7\times10^{-6})$ 
and
${\cal P}^b_{\tilde g\tilde g} \approx 4.3\times10^{-5}\;(6.6\times10^{-6})$,
respectively. 
Around a 20\% of this probability corresponds to the
primary proton, whereas the rest expresses the probability
that the gluino pair is produced by secondary hadrons inside
the shower. Notice that the shower with no leading hadron
after the first interaction has more hadrons of intermediate
and lower energies. For a light gluino these tend to compensate
the effect of the leading hadron 
({\it i.e.}, ${\cal P}^b_{\tilde g\tilde g} \approx
{\cal P}^a_{\tilde g\tilde g}$). However, as the gluino mass $M$
increases the less energetic hadrons are not able to produce
gluinos and ${\cal P}^b_{\tilde g\tilde g}$ becomes significantly
smaller than ${\cal P}^a_{\tilde g\tilde g}$.

The probability to produce the pair of WIMPs in these two
showers is ${\cal P}^a_{\chi\chi}\approx 2.5\times10^{-8}$ and 
${\cal P}^b_{\chi\chi}\approx 3.1\times10^{-8}$. In this case
the exotic particles, of mass $M_\chi=70$ GeV, can be effectively
produced by hadrons of intermediate energy and 
${\cal P}^b_{\chi\chi}>{\cal P}^a_{\chi\chi}$.

\begin{figure}
\centerline{\includegraphics[width=\linewidth]{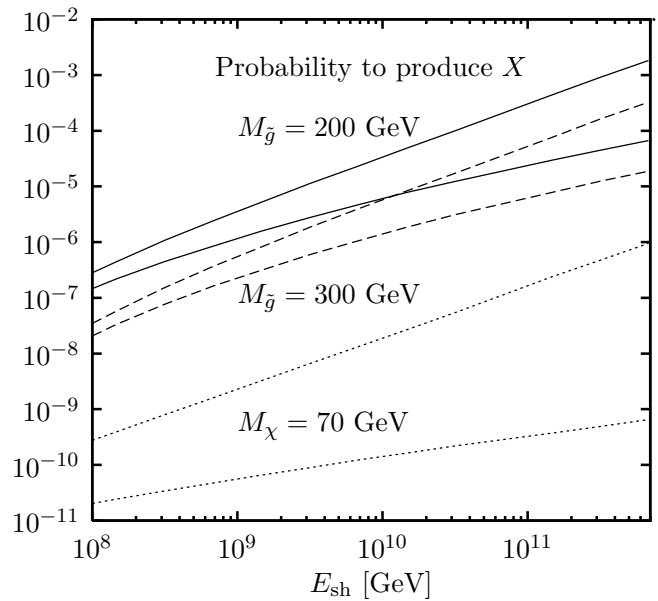}}
\caption{
Probability to create $X=\tilde g\tilde g,\ \chi\chi$ by a primary 
nucleon (lower lines) or by the primary or any secondary hadron inside
the shower (upper lines) as a function of the shower energy. \label{fig6}}
\end{figure}

In Fig.~\ref{fig6} we plot the average probability to produce gluinos
(and WIMPs)
for a primary proton of energy above $10^8$ GeV.
We separate the probability that the gluinos are produced by
the primary or secondary hadrons inside the 
shower. We find that secondary pions are the main source
of gluinos in showers above $10^8$~GeV.

%%%%%%%%%%%%%%%%%%%%%%%%%%%%%%%%%%%%%%%%%%%%%%%%%%%%%%%%%%%%%%%%%%%%%%%%%%%%%%%
%%%%%%%%%%%%%%%%%%%%%%%%%%%%%%%%%%%%%%%%%%%%%%%%%%%%%%%%%%%%%%%%%%%%%%%%%%%%%%%

\section{Detectability of the gluino pairs}

The small flux of WIMPs that we have obtained
does not seem to imply any experimental consequences
in neutrino telescopes or air shower experiments.  
The flux of gluino pairs, however, might be visible.

As mentioned above, IceCube (around 1 km$^2$ of area) \cite{icecube}
would be exposed to one gluino pair
per year if $M\approx 160$~GeV. Since this mass is already below
the experimental limits from the Tevatron, such 
an event would be {\it unexpected}. 
For example, if $M=200\;(300)$~GeV
the probability to have the two gluinos crossing IceCube in one 
year is just around 0.22 (0.015). Let us assume that 
such unexpected event actually happens. The mean energy 
of the gluino pair is around
$E_0= 2.1\times10^6\;(5.1\times10^6)$~GeV, and we obtain an 
average angle between the two gluinos 
$\theta_{\tilde g\tilde g}\approx 8.8\times10^{-4}\;(5.4\times10^{-4})$~rad. 
We will approximate a constant interaction 
length in water $\lambda_{\tilde G}\approx 175$~g/cm$^2$ for the
gluino hadron and 
a linear energy loss per interaction $\Delta E=k \gamma$ with
$k\approx 0.2$~GeV. This means that a $\tilde G$ of $M=200$~GeV 
deposits a 1 per mille of its energy every 1.8 meters of water.
It is then easy to deduce that the energy of the gluino pair
at a depth $x$ is just
\beq
E_{\tilde G}\approx E_0\; \exp\{-x/x_0\}
\;,
\eeq
where 
$x_0=\lambda_{\tilde G}M/k\approx 1.8\times10^5\;(2.6\times10^5)$~g/cm$^2$ 
for 
$M=200\;(300)$~GeV. Therefore, two vertical gluinos would reach a depth
$x_1=1.4\times 10^5$ g/cm$^2$ (the top of IceCube) with a total 
energy around $E_1=1.1\times10^6\;(2.1\times10^6)$~GeV. 
If they are produced at an altitude $H=20$ km, the gluinos
will enter IceCube
separated by 18 (11) meters and will deposit 
$\Delta E=4.7\times10^5\;(1.5\times10^5)$~GeV of energy
through the kilometer long detector. Such a
signal could be clearly distinguished from a typical muon
bundle in an air shower core.

\begin{figure}
\centerline{\includegraphics[width=\linewidth]{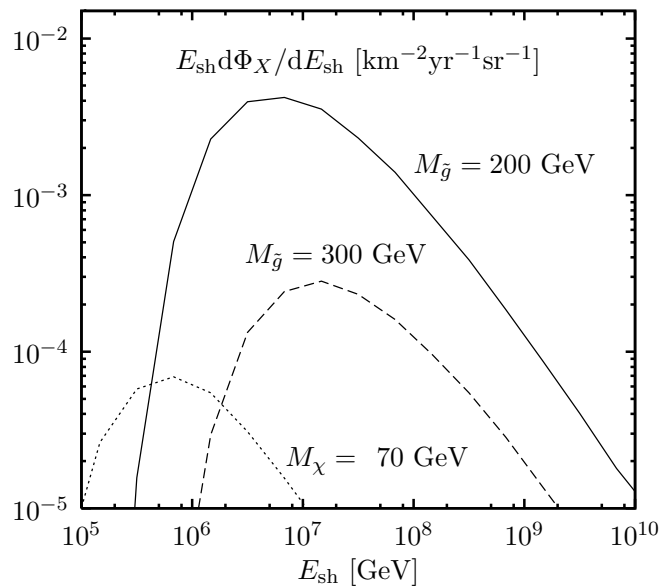}}
\caption{
Distribution of the gluino pairs as a function of the shower energy. 
\label{fig7}}
\end{figure}

More frequent events could be obtained
in the {\it larger} air shower experiment AUGER. The number
of gluino pairs of $M=200\; (300)$~GeV hitting per year
the $3000$~km$^2$ AUGER area would be around 
\beq
N_{\tilde g\tilde g}\approx \int_0^1 {\rm d}\cos\theta \; 
2\pi A \cos\theta \;T\; \Phi_{gg}\approx 330\;(22)
\;,
\eeq
where $\theta$ is the zenith angle. However, this is not a very
significant number, as these gluinos are {\it inside} the air
shower where they were produced, and many of these showers
have an energy below the $10^8$~GeV threshold in AUGER.
In Fig.~\ref{fig7} we plot the distribution of the 330 (22) gluino 
pairs as a function of the primary energy.
We obtain an estimate of 20 (2) gluino events 
per year inside showers above $10^8$~GeV.
A 75\% of these events come from 
a zenith angle below $60^\circ$ whereas the remaining 25\% are 
showers reaching AUGER quasi-horizontally, with a larger
zenith angle. 

%%%%%%%%%%%%%%%%%%%%%%%%%%%%%%%%%%%%%%%%%%%%%%%%%%%%%%%%%%%%%%%%%%%%%%%%%%%%%%%
%%%%%%%%%%%%%%%%%%%%%%%%%%%%%%%%%%%%%%%%%%%%%%%%%%%%%%%%%%%%%%%%%%%%%%%%%%%%%%%

Let us then analyze a typical gluino event inside a $10^{10}$~GeV
shower for $M=200$~GeV. 
Once created, the average angle between the gluinos is
$4.7\times10^{-4}$ rad, whereas the total energy of 
the pair is around $5.2\times10^7$~GeV.
The gluino-gluon R-hadrons will interact every 160 g/cm$^2$, depositing 
in each interaction an average one per mille of their energy 
(this corresponds to the value $k=0.2$ given above). Notice that the 
signal produced by a gluino hadron is quite different from that
produced by a light hadron of the same energy 
\cite{Anchordoqui:2004bd,Gonzalez:2005bc}. A proton 
deposits most of its energy within 
a couple of vertical atmospheres (2000 g/cm$^2$), whereas the gluino
would interact in the same interval around 12 times, 
depositing a fraction $10^{-3}$ of hadronic energy every time.
The gluino signal 
is more {\it homogeneous}, after 4 or 5 interaction lengths
it is a trace of constant energy. Notice also that the number of muons 
from pion decays that the gluino produces grows proportional to 
the length of that trace. This will affect the curvature 
in the shower front hitting the ground. 
The series of {\it mini-showers} started by the gluinos will 
produce muon bundles at different slant-depths, and the 
front will exhibit a more pronounced
curvature than what one would expect for a shower of the same
energy interacting only in the upper atmosphere.
For isolated gluino showers ({\it i.e.,} not {\it inside} a 
proton event) these features have been discussed 
by Anchordoqui {\it et al.} \cite{Anchordoqui:2004bd} and 
confirmed with a 
Montecarlo simulation in \cite{Gonzalez:2005bc}.
Previous works on air showers started by (lighter) long-lived 
gluinos were proposed as a possible explanation of the
events beyond the GZK cutoff \cite{Albuquerque:1998va}.

Therefore, one needs to 
distinguish two types of events, the {\it vertical}
ones (with a zenith angle $\theta < 60^\circ$) and the 
{\it quasi-horizontal} ones (with $\theta > 60^\circ$). If the
gluinos are produced in a  
$10^{10}$~GeV vertical shower they will take a 0.5\% 
of the initial energy and deposit a 6\% of that energy in two
traces separated by $\approx 9$ m. Each trace is the superposition
of hadronic showers of $2.6\times10^4$~GeV separated by 160~g/cm$^2$, 
{\it i.e.}, around $1.6\times 10^5$~GeV of energy in each gluino trace.
It seems unlikely that such a distortion in the profile of a 
vertical shower can be detected in AUGER.
Quasi-horizontal showers, however, could provide 
a more promising signal.
At large zenith angles the depth of the atmosphere increases
up to 36000 g/cm$^2$ for $\theta=\pi/2$,
and this has several important effects on the gluino signal. First,
the two gluino hadrons will cross the AUGER surface with a larger 
separation than in vertical showers. 
In particular, horizontal events come from a distance
$D \approx \sqrt {2 H R_T}$, where $H\approx 20$ km and $R_T$ 
is the radius of the
Earth. This gives a separation $\approx D \theta_{\tilde g\tilde g}$.
Second, 
after a certain depth most of the hadronic energy of the shower 
has been absorbed
by the atmosphere, leaving only muons (and invisible neutrinos).
In contrast, the hadron content in the gluino trace is basically
constant, as each $\tilde G$ starts a $2.6\times10^4$ GeV shower every 160
g/cm$^2$. Therefore, the detectors would observe charged
hadrons produced by the two gluinos in addition to the muons. 
The third effect on these inclined
events is that the total number of muons produced in the gluino traces
is larger than in vertical showers,
as it grows almost linearly with the depth. In this
way, the two gluinos inside the $10^{10}$ GeV event would deposit
around $10^7$ GeV (a $20\%$ of their total energy) if they come horizontally
crossing a depth of 36000 g/cm$^2$. Notice that these muons produced
deep in the atmosphere increase the curvature of the
shower front. 

\section{Summary and discussion}

Cosmic ray experiments could provide a window to explore
physics beyond the 
standard model, complementing accelerator experiments. 
The primary nucleons or secondary hadrons inside an air shower
may produce new massive particles when they interact
with atmospheric nucleons. We first have determined the total
flux of hadrons within cosmic rays. We think Fig.~\ref{fig1} is a
relevant result, as it is this (and not just the flux of 
primaries) the relevant flux to evaluate the 
production rate of new physics by cosmic rays. Then we
have discussed how to calculate the creation of a 
massive particle, focussing on the long-lived gluino
of split-SUSY models. We have shown that 
the flux of gluino hadrons produced
by very energetic cosmic rays allows the possibility of
isolated gluino events crossing a km$^2$ neutrino telescope 
like IceCube. In the larger detector AUGER, for a gluino
mass $M=200$~GeV we obtain an
estimate of 20 gluino events per year inside air showers of energy
above $10^8$~GeV. We have argued that the gluinos could
modify substantially the profile of the showers with 
a large zenith angle. The gluino hadrons propagate
losing a 1 per mille of their energy in every interaction length,
defining an air shower with an approximately constant number 
of hadrons and with a number of muons that grows linearly with
the depth. Therefore, the signal is stronger if they are
created inside showers that 
reach AUGER from a large zenith angle. 
We think that the possibility of detecting in AUGER such a signal  
deserves a more detailed study.

We thank Carlos Garc\'\i a-Canal and Paolo Lipari for useful 
discussions. This work has been supported by MEC of Spain 
(FPA2003-09298-C02-01 and FPA2006-05294) and Junta de Andaluc\'\i a 
(FQM-101 and FQM 437). 
We also acknowledge finantial support from a grant CICYT-INFN (06-13).

\appendix

\section{Partonic cross sections}

Making use of the Mandelstam variables, $s+t+u=2\mgsq$, and the polar angle $\theta$ in the center of mass frame, $t=-(1+\beta^2-2\beta\cos\theta)s/4$, with $\beta=\sqrt{1-4\mgsq/s}$ and $\gamma=(1-\beta^2)^{-1/2}$, the cross-sections read:
\beq
\frac{{\rm d}\sigma}{{\rm d}\cos\theta}=\frac{s\beta}{2}\frac{{\rm d}\sigma}{{\rm d}t},\quad
\sigma=\frac{1}{2}\int_{-1}^1{\rm d}\cos\theta\ \frac{{\rm d}\sigma}{{\rm d}\cos\theta}.
\eeq

\noindent
\underline{$gg\to\tilde g\tilde g$:}
\beqa
\frac{{\rm d}\sigma}{{\rm d}t}&=&\frac{9\pi\alpha_s^2}{4s^2}
\left\{
 \frac{2(t-\mgsq)(u-\mgsq)}{s^2}
+\frac{\mgsq(s-4\mgsq)}{(t-\mgsq)(u-\mgsq)}
\right.
\nonumber \\
&&+\frac{(t-\mgsq)(u-\mgsq)-2\mgsq(t+\mgsq)}{(t-\mgsq)^2}
\nonumber \\
&&+\frac{(t-\mgsq)(u-\mgsq)+\mgsq(u-t)}{s(t-\mgsq)}
\nonumber \\
&&
+\frac{(u-\mgsq)(t-\mgsq)-2\mgsq(u+\mgsq)}{(u-\mgsq)^2}
\nonumber \\
&&\left.+\frac{(u-\mgsq)(t-\mgsq)+\mgsq(t-u)}{s(u-\mgsq)}
\right\},
\\
\sigma&=&\frac{3\pi\alpha_s^2}{4s}\left\{
3\left(1+\frac{1}{\gamma}
-\frac{1}{4\gamma^2}\right)\log\frac{1+\beta}{1-\beta}
\right.
\nonumber \\
&&
\left.-\left(4+\frac{17}{2\gamma^2}\right)\beta
\right\}.
\eeqa

\noindent
\underline{$q\bar q\to\tilde g\tilde g$:}
\beqa
\frac{{\rm d}\sigma}{{\rm d}t}&=&\frac{8\pi\alpha_s^2}{3s^2}
\frac{(t-\mgsq)^2+(u-\mgsq)^2+2\mgsq s}{s^2},
\\
\sigma&=&\frac{8\pi\alpha_s^2}{9s}
\left(1+\frac{1}{\gamma}\right)\beta.
\eeqa

\noindent
\underline{$q\bar q\to\chi\chi$:}
\beqa
\frac{{\rm d}\sigma}{{\rm d}\cos\theta}&=&
\frac{\pi\alpha^2\epsilon^2}{32s^4_Wc^4_W}\frac{s\beta^3}{(s-M_Z^2)^2}
\\&&\times
(1-4\eta+8\eta^2)(1+\cos^2\theta),
\\
\sigma&=&
\frac{\pi\alpha^2\epsilon^2}{24s^4_Wc^4_W}\frac{s\beta^3}{(s-M_Z^2)^2}
(1-4\eta+8\eta^2),
\eeqa
with $\eta=|Q_q|s^2_W$ and $\epsilon=0.25$ (see discussion in Sect.~III).

\end{document}